\documentclass[onecolumn]{aa}
\usepackage{graphicx}

\begin{document}

\title{The nature of the ultraluminous X-ray sources inside galaxies and
their relation to local QSOs}

\author{G. Burbidge
\inst{1}
\and E. M. Burbidge
\inst{1}
\and H. Arp
\inst{2}}

\offprints {G. Burbidge}

\institute{Center for Astrophysics and Space Sciences
0424, University of California, San Diego, CA 92093-0424, USA\\
\email{gburbidge@ucsd.edu}
\and
Max-Planck-Institut f\"ur Astrophysik, Karl Schwarzschild-Str.1,
  Postfach 1317, D-85741 Garching, Germany\\
   \email{arp@mpa-garching.mpg.de}}

\date{Received}

\abstract{
It is suggested that many of the
ultraluminous compact x-ray sources now being found in the main
bodies of galaxies, particularly those that are active, like M82,
NGC 3628 and others, are``local" QSOs, or BL Lac objects, with
high intrinsic redshifts in the process of being ejected
from those galaxies.  Evidence in support of this hypothesis is summarized. 
}

\titlerunning{ultraluminous x-ray sources}

\maketitle

\section{Introduction} 

It is generally agreed that there are two types of compact x-ray
sources whose properties are well established. The first objects
of this kind are the x-ray sources that arise in stellar binary
systems.  These are clearly associated with accretion disks around
highly evolved stars - neutron stars and black holes. Since there
is much evidence that the upper end of the stellar mass function
is close to 100 M{$_\odot$}, it is clear that there is a
luminosity limit of 10$^{39}$ erg sec$^{-1}$ that can
be expected from any x-ray
binary system which has arisen from a normal stellar system. \\

The second type of compact x-ray source is that
associated with active galactic nuclei and quasi-stellar objects.
These sources have much greater luminosities, which in the
conventional view are attributed to the presence of much more
massive black holes in the
centers of galaxies and in QSOs.\\

In studies of comparatively nearby galaxies ROSAT began to find
nuclear sources with L ${\simeq} 10^{39} - 10^{40}$ erg
sec$^{-1}$, corresponding to black hole accretion sources with
masses in the range 10$^{3} - 10^{4}$ M${_\odot}$.  However, in
recent years, a number of studies have shown that x-ray sources
with this range of luminosity are often present in the main bodies
of spiral and irregular galaxies and not just in the nuclei. These
discoveries, due to the high resolution properties of Chandra and
XMM Newton, show that ultraluminous sources are present in NGC
3628, 4038-39, 4565, 4698, M82, NGC 5204 and other galaxies
(Strickland et al 2001; Fabbiano et al 2001; Foschini et al 2002a;
Kaaret et al 2001, Makishima et al 2000, Wu et al. 2002). A recent
study of a sample of nearby Seyfert galaxies with XMM-Newton has
shown that they also contain many off-center compact x-ray sources
with luminosities in the range 10$^{39} - 10^{40}$ erg sec $^{-1}$
(Foschini et al 2002b). \\

Several proposals to explain these sources have been made, the
main one being the suggestion that intermediate size black holes
with masses 10$^{2} - 10^{4}$ M${_\odot}$ are responsible and the
energy release is again through accretion. Alternatively it has
been proposed that some of these sources are background QSOs or BL
Lac objects. However, for those which lie very close to the nuclei
of the galaxies the chance that they are background sources is
very small. \\

In this note we propose that these sources are (local) low
luminosity QSOs which are in the process of being ejected from
their places of birth in the nuclei of the galaxies.  In the
following section we summarize evidence in support of this
hypothesis. \\

\section {Evidence for clustering of QSOs
about galaxies} 

Following the early work based on statistical evidence for the
physical association between nearby galaxies and bright high
redshift QSOs, and individual examples, which has been extensively
summarized elsewhere (cf Burbidge et al 1971, Arp 1967, 1987,
Burbidge 2001, Hoyle et al 2000) there have been a number of
recent investigations which strongly suggest that QSOs can be
detected as they are ejected from low redshift active galaxies.
Some of these are as follows:

\begin{itemize}

\item Two compact x-ray QSOs ejected symmetrically along an axis of
NGC 4258 (distance 7 Mpc) (Pietsch et al 1994; Burbidge 1995);

\item Two compact x-ray emitting QSOs apparently ejected from NGC 2639
(Burbidge 1997);

\item Two pairs of x-ray emitting QSOs, one pair
with almost equal values of the redshift (z = 1.25 and 1.26)
associated with Arp 220 (Arp, Burbidge et al 2001);

\item A number of QSOs with morphological connections and a
configuration strongly suggesting recent ejection in NGC 3628 with
some QSOs exceedingly close to the nucleus of the galaxy (Arp, Burbidge et al 2002).

\item Five QSOs aligned along an axis associated with NGC 3516 (Chu et
al 1998). \\
\end{itemize}

These are only some of the many cases which have now been found
which provide strong circumstantial evidence that QSOs originate
in the centers of active galaxies and are ejected from them. \\

Evidence based on a sample of 39 x-ray emitting QSOs associated
with active galaxies, including some of the cases listed above,
suggests that the line of sight components of the velocities of
ejection average about 12000 km/sec$^{-1}$ (both redshifted and
blueshifted) (cf Burbidge and Napier 2001).  Of course the QSOs,
since they are at the distances of the galaxies and have typical
apparent magnitudes in the range 17$^{m} - 20^{m}$, are much
fainter than the galaxies.  Thus at the distance of NGC 4258 (7
Mpc) the absolute magnitudes of the ejected QSOs are only about M
= -10; i.e. they are no brighter than the brightest O stars.
However, the detected QSOs associated with Arp 220, using its
redshift z = 0.018 and H${_0}$ = 60 km sec $^{-1}$ Mpc$^{-1}$ lie
at a distance of $\sim$ 90 Mpc.
Thus they are much more luminous with M = -16. \\

In the cases in which QSOs have been found to be clustered about
active galaxies the scale of the clustering is $\sim 5^{\prime} - 50^{\prime}$.
Thus it is clear that there will be an effect that must be taken into
account, associated with the fact that since there is a
considerable range in the distances of the parent galaxies, we
shall identify intrinsically faint QSOs close to nearby galaxies,
but only the brighter QSOs will be seen to be associated with
parent galaxies which are much farther away.  For the more distant
parent galaxies, the intrinsically fainter QSOs will not be
detected. \\

For example, if NGC 4258 was at the same distance as Arp 220, the
two QSOs would have m $\simeq$ 24, and they would lie less than 40
arc seconds from the center of the galaxy.  Thus they would not be
detected. Correspondingly, if QSOs as bright as those detected around
Arp 220 have been ejected from NGC 4258 they will lie at angular distances
of 1-2 degrees and thus they will not be thought to be associated with the galaxy.\\

In general, we suppose that dispersing clouds of QSOs with a wide
range of luminosities exist around most active galaxies, and they
range in luminosity in x-rays and/or optical flux from values $\sim
10^{39}$ erg sec$^{-1}$ upward.  Since they are born in the nucleus, and
eventually are found outside, it is reasonable to suppose that
some of them will be detected in the main body of the galaxy, as
they travel out. Thus it is clear also that many of them will be
found inside the main luminous body of the galaxy.  In our view
these are the ultraluminous x-ray sources that are beginning to be
found in comparatively close by galaxies. \\

One piece of evidence which tends to bear this out is associated
with M82. In this galaxy a number of QSOs have been found very
close to the main body.  M82 is the nearest active galaxy (D =
3.25 Mpc).  We have recently discovered many x-ray QSOs in
addition to those which were discovered earlier in this system
(Burbidge et al. 2002).  Most of the QSOs lie within 10$^{\prime}$
or less of the center of M82 and they are quite faint $\sim 18^{m}
- 20^{m}$. This means that they have absolute magnitudes in the
range $-7.4 < M < -9.4$. The configuration of the QSOs about M82
strongly suggests that they are associated with it.  We therefore
believe that the x-ray sources found in the main body of M82 by
Kaaret et al (2001) and Makishima et al (2000) are also QSOs in
the process of ejection. If M82 was observed at the distance of
Arp 220, all of the QSOs which we have found to be associated with
it would be much too faint, and much too close to the main body of
the galaxy to be detectable.  However, they will contribute
significantly to its
x-ray luminosity. \\

Obviously, the test of this general hypothesis is to detect a QSO
in the main body of a galaxy and show that it has a large
redshift.  Has this already been achieved? \\

\section{Sources in NGC 5204 and NGC 4698}

\noindent \underline{NGC 5204}

Roberts et al (2001) have recently found an optical counterpart to
an x-ray source in NGC 5204 which they call NGC 5204 X-1.  At the
distance of NGC 5204, it has an x-ray luminosity of 5.2 x
10$^{39}$ erg sec$^{-1}$.  It has a blue continuum with m${_v}$ =
19.7.  At the distance of NGC 5204 M${_v}$ = 8.7. Roberts et al
point out that the spectrum is featureless in all wavelengths,
point like, and it shows long term x-ray variability. Thus they
suggest that it may be a background BL Lac object.  We suggest
that it is a genuine BL object within the body
of NGC 5204 in the process of being ejected from it. \\

\noindent \underline{NGC 4698}

Foschini et al (2002) have reported that they have found
what they call a background BL Lac object with z = 0.43 in NGC
4698 (z = 0.0033).  This clearly can be interpreted as evidence
for a local QSO with an intrinsic redshift of 0.43 being ejected
from NGC 4698.  The object has an x-ray luminosity of 3 x 10$^{39}$
ergs sec$^{-1}$. \\

\section{Conclusion}

There is circumstantial evidence suggesting that some of the
recently discovered ultraluminous x-ray sources in the bodies of
comparatively nearby galaxies are local QSOs and BL Lac objects
being ejected from those galaxies. \\

This evidence simply comes from the fact that many comparatively
nearby galaxies in different stages of activity are surrounded by
``local'' QSOs which are physically associated with them.  The
obvious conclusion is that these QSOs were created in the nuclei
of the galaxies, and have been ejected from them.  The speeds of
ejection are apparently $\leq$ 0.1c.  Thus, such QSOs will take
$\geq$ 10$^{6}$ years to escape from the main body of a galaxy.
Consequently we would expect to find some of them inside the
galaxy, moving
outward. \\

\noindent
Tests of this hypothesis are:

\begin{itemize}
\item To obtain optical spectra of such objects and show that they
have large redshifts.  This may have already been accomplished in
the case of the QSO in NGC 4698.

\item To attempt to detect proper motions of such sources.  For
sources in M82, a speed of 0.1c across the line of sight, which may be
high, corresponds to a proper motion of about 2 milliarc seconds
per year.
\end{itemize}

Of course, this explanation does not exclude the possibility that
some of these sources are indeed accreting objects containing
black holes with M = $10^{3} - 10^{4} M_{\odot}$.  Such sources
will have the normal properties of accreting objects, and will be
distinguishable from local QSOs because they will not have large
redshifts or comparatively large proper motions. \\

In all of the cases where there is circumstantial evidence that
QSOs are being ejected from an active galaxy the parent galaxy is
a spiral or irregular system.  In the case of NGC 4697 which is an
elliptical galaxy, many luminous x-ray sources have also been
detected outside the nucleus (Sarazin et al 2001).  The
luminosities range from 5 x 10$^{37}$ erg sec$^{-1}$ to 2.5 x 10$^{39}$ 
erg sec$^{-1}$
and Sarazin et al have suggested that these arise in accreting
systems containing black holes with M = 10$^{3} - 10^{4} M_{\odot}$.  
We consider that this is probably correct. \\

In our opinion the universe is sufficiently complex for both types
of beasts to exist in it.  The distinction between them is that
the active variable objects will have intrinsic redshifts and will
only be found in galaxies with active nuclei or star-forming
activity, while the accreting massive black hole systems will
largely be thermal emitters and will tend to be found in
ellipticals. \\

\end{document}